\documentclass[reprint,pre,
twocolumn,
notitlepage,longbibliography]{revtex4-1}

\usepackage{graphicx} 
\usepackage{float}
\usepackage{array} 
\usepackage{paralist} 
\usepackage{subfigure} 

\usepackage{fancyhdr} 
\usepackage{threeparttablex}
\usepackage{amsmath}
\usepackage{tcolorbox}
\usepackage{algorithm}
\usepackage{tikz-inet}	
	\usetikzlibrary{graphs,trees,shapes,snakes,arrows,circuits.logic.US}
	\usepackage{circuitikz}
	\usetikzlibrary{backgrounds,fit,decorations.pathreplacing} 
	\tikzstyle{branch}=[fill,shape=circle,minimum size=3pt,inner sep=0pt]
	
	\usepackage{xcolor}
\newcommand{\ket}[1]{\ensuremath{\left|#1\right\rangle}} 
\newcommand{\bra}[1]{\ensuremath{\left\langle#1\right|}} 

\begin{document}
\title{Obtaining A Linear Combination of the Principal Components of a Matrix on Quantum Computers} 
 \author{Anmer~Daskin}\email{email: anmerdaskin@yahoo.com}
\affiliation{Department of Computer Engineering, Istanbul Medeniyet University, Kadikoy, Istanbul, Turkey}
\begin{abstract}

Principal component analysis  is a  multivariate  statistical  method  frequently used in science and engineering to reduce the dimension of a problem  or   extract the most significant features from a dataset. 
In this paper, using a similar notion to the quantum counting,  we  show how to apply the amplitude amplification together with the phase estimation algorithm  to an operator in order to  procure the eigenvectors of the operator associated to the eigenvalues defined in the range $\left[a, b\right]$, where $a$ and $b$ are real and $0 \leq  a \leq b \leq 1$. This makes possible to obtain a combination of the eigenvectors associated to the largest eigenvalues and so can be used to do principal component analysis on quantum computers. 
\end{abstract}

\maketitle
\section{Introduction}

Principal component analysis (PCA) is widely used as a multivariate statistical  technique  to reduce the dimension of a problem  or  to extract most important features from high dimensional datasets by fitting ellipsoids representing principal components to the given dataset \cite{Pearson1901, Jolliffe2002}.
Examples of PCA with different descriptions and settings include probabilistic PCA \cite{Tipping1999probabilistic}, nonlinear PCA \cite{De2005nonlinear}, kernel PCA \cite{Scholkopf1997kernel}  as a nonlinear extension of PCA, local PCA \cite{Kambhatla1997dimension} for dimension reduction, and  generalized PCA \cite{Vidal2005generalized}, applied to a wide variety of problems in science and engineering \cite{Jolliffe2002}: e.g., it is  a prevalent method  in pattern recognition \cite{Bishop2006pattern}.
Principal component analysis  best explains the variation in the whole dataset represented by an $I\times J$ matrix $X$  representing
$I$ observations described by
$J$ variables. 
 For a set of data vectors,  in PCA, one finds $q$ orthonormal axes  describing a linear projection of the dataset under which the retained variance is maximum. 
 These $q$ orthonormal axes-so called the principal components-corresponds to the $q$ dominant eigenvectors associated with the $q$ largest eigenvalues of the  covariance matrix.  
The projection onto the subspace spanned by these eigenvectors  not only eliminates the possible noise in the data but  reduces the size of the original high dimensional data as well \cite{Tipping1999probabilistic}.
To obtain the eigenvectors corresponding to the $q$ largest eigenvalues, generally singular value or eigenvalue decomposition techniques are applied to the matrix $X$ or to the covariance matrix $XX^T$. 
For unfamiliar readers, we recommend the simple introductory tutorial given in Ref. \cite{Shlens2014tutorial} and the book in Ref.\cite{Jolliffe2002} for further information on PCA and its applications.

 Eigenvalue related problems can be solved successfully on quantum computers by using the well-known quantum phase estimation algorithm (PEA) \cite{Kitaev1996}. 
 Given an input initial approximate eigenvector, PEA  generates the associated eigenvalue in a more efficient way than all the known classical algorithms \cite{Papageorgiou2013fast}. 
 However, finding a good approximate initial state required for the algorithm is a nontrivial task which inevitably impinges  on the efficiency. 
 PEA with an equal superposition state is shown to eliminate the need for an approximate initial eigenvector for obtaining the eigenvector associated with the largest eigenvalue of a stochastic matrix \cite{Daskin2014mna,Daskin2015ppi}. PEA is also applied to solve the  linear systems of equations given in the form $A\bf{x}=\bf{b}$. In that case, the initial state is mapped to the vector $\bf b$ and the solution $\bf x$ is obtained as the final state. In addition, it is shown that a flag register can be used to eliminate the ill-conditioned part of $A$ and process only eigenvalues greater than some threshold value \cite{Harrow2009quantum}. 
 
Research in quantum algorithms proposed for machine learning and analyses of big data  have gained  a lot of momentum in recent years (e.g. see \cite{Rebentrost2014quantum, Wittek2014quantum, Schuld2014quantum}). 
Speed-ups over the conventional classical methods are shown to be achievable mostly by using adiabatic quantum computation \cite{Neven2008training,Neven2009training} or mapping data set to  quantum random access memory \cite{Aimeur2013quantum, Lloyd2014topological}. Furthermore, in Ref.\cite{lloyd2014qQPCA}, Lloyd et al. have described a method for quantum principal component analysis: 
In their method, multiple copies of a quantum system defined by the low rank density matrix $\rho$ are used to obtain $e^{i\rho t}$ efficiently.
Then, the phase estimation algorithm with the initial state 
$\rho$ itself is applied to generate a final quantum state $\sum_i \lambda_i \ket{\varphi_i} \bra{\varphi_i}\otimes \ket{\hat{\lambda}_i} \bra{\hat{\lambda}_i}$, where  $\lambda_i$s, $\ket{\varphi_i}$s and $\hat{\lambda}_i$s are respectively the eigenvalues, the eigenvectors and the estimates of the eigenvalues of  $\rho$. Sampling from this final quantum state, one is obviously able to obtain any of the eigenvectors. In addition, when $\rho$ is a low rank matrix, they have showed that the final state is in superposition of the principal eigenvectors, which can be used in the principal component analysis.

In this paper, we present a different general framework to do principal component analysis on quantum computers by showing how to generate the eigenvalues defined in certain range and their corresponding eigenvectors in the phase estimation algorithm. In particular, as done in Ref. \cite{Daskin2014mna,Daskin2015ppi}, we  first use the phase estimation algorithm with an initial  superposition state to generate an output state which has the superposition of the eigenvalues of a given operator on the first register and the associated eigenvectors on the second register. We then show how to apply the amplitude amplification to this output state to amplify the amplitudes of the desired eigenvalues and their associated eigenvectors. We then give a numerical example and finally discuss the future research directions.

\section{Preliminaries}

\subsection{Quantum Phase Estimation Algorithm}
When there exists an  initial estimate for the eigenvector of a $2^n \times 2^n$ operator $\mathcal{H}$, PEA finds the associated eigenvalue by using the powers of the time evolution operator $U=e^{i2\pi \mathcal{H}}$. 
PEA consists of two registers: the first register, \ket{reg_1}, holds the eigenvalue at the end with the precision determined by the number of qubits, $m$, allocated for this register. 
 On the other hand, \ket{reg_2} with $n$ number of qubits is used for the eigenvector. 
The algorithm starts with the initial state \ket{reg_1}\ket{reg_2}=\ket{\bf 0}\ket{\varphi_j}, where \ket{\varphi_j} is the  eigenvector associated to $\lambda_j$, the $j$th eigenvalue of $\mathcal{H}$. 
It then applies $U_{QFT}^\dagger CU^{2^m} \dots CU^{2^0} U_{QFT}$ to the initial state. Here, $CU^{2^k}$ describes  a  quantum operation: controlled by the $k$th qubit in the first register and applied to the second register. And
$U_{QFT}$ and $U_{QFT}^\dagger$ are respectively the quantum Fourier and  inverse Fourier transforms applied to the first register.   The final quantum state holds the eigenvalue $\lambda_j$ on the first register and the associated eigenvector on the second register:
\begin{equation}
\ket{\lambda_j}\ket{\varphi_j} = 
U_{QFT}^\dagger CU^{2^{m-1}} \dots CU^{2^0} U_{QFT} \ket{\bf 0}\ket{\varphi_j}.
\end{equation} 
Therefore, measuring the first register produces $\lambda_j$ with the precision determined by the number of qubits in the first register.

\subsection{Quantum Amplitude Amplification}
Grover's search algorithm \cite{Grover1998} can be applied to amplify the amplitude of a solution encoded in a quantum state. 
Consider an $N\times N$ quantum operator (or an algorithm) $A$ mapping an initial quantum state \ket{\bf{0}} into \ket{\psi} \cite{Mosca1998quantum, Brassard2002,Kaye2006}:
\begin{equation}
\label{EqPsi1}
\begin{split}
\ket{\psi}= & A\ket{\bf 0}=\sum_{x=0}^N\alpha_x\ket{\bf x}\ket{\phi}
\\ =
& \sum_{x\in X_{good}}\alpha_x\ket{\bf x}\ket{\phi} + \sum_{x\in X_{bad}}\alpha_x\ket{\bf x}\ket{\phi},
\end{split}
\end{equation}
where $\alpha_x$s are the complex amplitudes,  $\ket{\bf x}$ represents the $x$th vector in the standard basis, and \ket{\phi} is considered as some ``junk" state or additional workspace (This  state shall be used for the eigenvector in the next section.).
 From this equation, the probabilities to see the ``good" and the ``bad"  states described by the sets $X_{good}$  and $X_{bad}$, respectively, are as follows:
\begin{equation}
\begin{split}
&P_{good} = \sum_{x\in X_{good}}|\alpha_x|^2 
\text{\ and\ } \\
& P_{bad} = \sum_{x\in X_{bad}}|\alpha_x|^2.
\end{split}
\end{equation}
When these probabilities inside (0,1), i.e. $0<P_{good} <1$, an orthonormal basis set for two dimensional subspace can be formed by  separating and renormalizing the ``good" and  the ``bad" components of the state \ket{\psi} in Eq.(\ref{EqPsi1}):
\begin{equation}
\begin{split}
& \ket{\psi_{good}}=  \sum_{x\in X_{good}}\frac{\alpha_x}{\sqrt{P_{good}}}\ket{\bf x}\ket{\phi} \text{\ and \ }
\\ &\ket{\psi_{bad}}
= 
\sum_{x\in X_{bad}}\frac{\alpha_x}{\sqrt{P_{bad}}}\ket{\bf x}\ket{\phi}.
\end{split}
\end{equation}
In terms of the orthonormal basis $\{\ket{\psi_{good}}, \ket{\psi_{bad}}\}$, Eq.(\ref{EqPsi1}) can be rewritten as:
\begin{equation}
\ket{\psi}= sin(\theta)\ket{\psi_{good}}+cos(\theta)\ket{\psi_{bad}},
\end{equation}
where $sin(\theta)=\sqrt{P_{good}} = \sum_{x\in X_{good}}|\alpha_x|^2$ for $\theta \in [0,\frac{\pi}{2}]$.

It is known that to amplify the amplitude of the searched item, the  search iterate described by the following matrix product can be used:
\begin{equation}
\label{EqUpsiUf}
Q=AU_0^\perp A^{-1}U_f=U_{\psi}^{\perp} U_f.
\end{equation}
When the operators in the above applied to \ket{\psi}, they acts as follows: 
$Uf$ negates the amplitude of the searched item (the ``good" states) and does nothing for the other items (the ``bad" states):
\begin{equation}
U_f\ket{\psi}=-sin(\theta)\ket{\psi_{good}}+cos(\theta)\ket{\psi_{bad}}.
\end{equation}
In addition, the operator $U_f$ can be constructed by considering it as the reflection about the state $\ket{\psi_{good}}$: 
\begin{equation}
U_f=I-2\ket{\psi_{good}}\bra{\psi_{good}},
\end{equation}
where $I$ is the identity matrix.
  $U_0^\perp$ in Eq.(\ref{EqUpsiUf}) is the reflection about the axis defined by the vector \ket{\bf 0}:  $U_0^\perp=2\ket{\bf 0}\bra{\bf 0}-I$. And so the operator  $U_{\psi}^{\perp}$ in Eq.(\ref{EqUpsiUf}) reads the following:
\begin{equation} 
 U_{\psi}^{\perp}=2\ket{\psi}\bra{\psi}-I.
 \end{equation} 
 Note that this is also a reflection operator: when $U_{\psi}^{\perp}$ is applied to $\ket{\psi}$, it does nothing. However, it changes the sign of the states, \ket{\psi^\perp}, orthogonal to $\ket{\psi}$:
\begin{equation}
U_{\psi}^{\perp} \left( sin(\omega)\ket{\psi}+cos(\omega)\ket{\psi^\perp}
\right)=sin(\omega)\ket{\psi}-cos(\omega)\ket{\psi^\perp}
\end{equation}

Consequently, the  application of $Q=U_{\psi}^{\perp}U_f$  to \ket{\psi} rotates the state in the following way:
\begin{equation}
U_{\psi}^{\perp}U_f\ket{\psi} =sin(3\theta)\ket{\psi_{good}}+cos(3\theta)\ket{\psi_{bad}}
\end{equation}
For $0<\theta \leq \pi/6$, this obviously increases the magnitude of the amplitude of the state $\ket{\psi_{good}}$.  For  the $k$ number of  repeated applications of $Q$, the final state reads:
\begin{equation}
Q^k\ket{\psi} =
 sin\left((2k+1)\theta\right)\ket{\psi_{good}}
+\cos\left((2k+1)\theta\right)\ket{\psi_{bad}}.
\end{equation}
The highest probability occurs when $(2k+1)\theta \approx\frac{\pi}{2}$, which allows $k$ to be bounded by $\Omega(\frac{1}{\theta})$.
Since  $sin(\theta)=\theta$ when $\theta$ is small, we can conclude that the required number of iterations for the maximum probability is approximately $\frac{\pi}{4\theta}$, i.e., $k\approx  \frac{\pi}{4\theta}$.

\section{Quantum Principal Component Analysis}
In this section, we shall show how to use PEA and amplitude amplification to obtain the eigenvectors corresponding to the largest eigenvalues of an operator.

Consider that the eigenvalues of the operator $\mathcal{H}$ are ordered as $\lambda_0\geq \dots \geq\lambda_{n-1}$.
In the principal component analysis,  we deal with the eigenvectors associated with the  first $q$ principal eigenvalues. 
Accordingly, in the phase estimation algorithm, we need to find a way to produce a quantum state in which the first register is the superposition of the first $q$ eigenvalues while the second register represents the superposition of the associated eigenvectors.   
When the initial state in the phase estimation algorithm is in a superposition state, PEA generates the superposition of the eigenvectors and the eigenvalues with the amplitudes defined by the overlap of the eigenvectors with the initial superposition state. 

To get a quantum state which is the superposition state of these eigenvectors,  we first define a range $[a, b]$ with $0\leq a\leq b \leq 1$ expected to include the $q$ principal eigenvalues. 
Here, one can define a controlled measurement operator to get the state in the second register where the value of the first register falls in the range $\left[ a, b \right]$.
 However,  in the cases when the initial state is almost orthogonal to the $q$ eigenvectors; the success probability gets very low since it depends on the overlap of the eigenvectors and the initial state.

Instead of a controlled measurement; following a similar notion to  the well-known quantum counting \cite{Mosca2001139}, here, we apply the amplitude amplification to the output of the phase estimation algorithm: first, the eigenvalues lying in this range are marked and then, their amplitudes are amplified. In this amplification process, two operators are needed: one is to mark the desired amplitudes and the other one is to amplify the amplitude of the marked item. 
In the following subsection, we describe this process in detail in two parts: viz, the phase estimation part and the amplitude amplification part.

\subsection{Details of the Method}
\subsubsection{The Phase Estimation Part}
The first part of the method follows the standard phase estimation algorithm with the initial equal superposition state. The transition of the states in this part is as follows:
\begin{tcolorbox}
\begin{itemize}
\item Initialize both of the registers as
\begin{equation}
\ket{\psi_0}=\ket{reg_1}\ket{reg_2}= \ket{\bf 0}\ket{\bf 0}
\end{equation}
\item Apply the quantum Fourier and the Hadamard transforms to the first and the second registers, respectively:
\begin{equation}
\begin{split}
\ket{\psi_1}= & \left(U_{QFT}\otimes H^{\otimes^n}\right) \ket{\psi_0}
\\
=&
\frac{1}{\sqrt{2^m}}\sum_{j=0}^{2^m-1}\ket{\bf j}\otimes \frac{1}{\sqrt{2^n}}\sum_{j=0}^{2^n-1}\ket{\bf j}
\end{split}
\end{equation}
Note that, here, $U_{QFT}=H^{\otimes^n}$.
\item Apply the controlled Unitary operations to the second register controlled by the qubits in the first register:
\begin{equation}
\ket{\psi_2}=CU^{2^{m-1}}\dots CU^{2^0}\ket{\psi_1}= \sum_{j=0}^{2^n-1}\alpha_j e^{i\lambda_j}\ket{\bf j}\ket{\varphi_j},
\end{equation}
where $\alpha_j$ is the overlap between the initial state and the $j$th eigenstate: $\alpha_j=\sum_{k=1}^{2^n}
\langle {\bf{k}}, \varphi_j\rangle  $, which is the normalized sum of the vector elements of \ket{\varphi_j} when the initial state is the equal superposition state.
\item Apply the inverse quantum Fourier transform to the first register so as to obtain the superposition of the eigenvalues on this register:
\begin{equation}
\ket{\psi_3} = 
\left(U_{QFT^\dagger}\otimes I^{\otimes^n}\right)\ket{\psi_2}
=
\sum_{j=0}^{2^n-1}\alpha_j \ket{\lambda_j}\ket{\varphi_j},
\end{equation}
\end{itemize}
\end{tcolorbox}

\subsubsection{The Amplitude Amplification Part}
The second part of the method is the amplitude amplification part where the desired eigenvalues are first marked by the operator $U_f$, and then the amplitudes corresponding to those eigenvalues are amplified by the application of the operator $U_{\psi_3}^\perp$. 
While  $U_f$ is applied only to the first register,  $U_{\psi_3}^\perp$ is applied to the whole system.  This is summarized as follows:
\begin{tcolorbox}
\begin{itemize}
\item Apply the operator $U_f$ to mark the certain eigenvalues for the amlitude amplification:
\begin{equation}
\begin{split}
\ket{\psi_4} = & 
\left(U_f\otimes I^{\otimes^n}\right) \ket{\psi_3}
  \\ = & 
\sum_{j=0 \& j \neq x }^{2^n-1}\alpha_j \ket{ \lambda_j}\ket{\varphi_j} - \alpha_x \ket{\lambda_x}\ket{\varphi_x},
\end{split}
\end{equation}
where $\lambda_x$ is the marked eigenvalue.
\item Apply the second operator of the amplitude amplification defined as: 
\begin{equation}
U_{\psi_3}^\perp
= 
2\ket{\psi_3}\bra{\psi_3}-I^{\otimes^{n+m}} 
=
U_{PEA}U_{0}^\perp U_{PEA}^\dagger,
\end{equation}
with 
\begin{equation}
\begin{split}
U_{PEA} = & \left(U_{QFT^\dagger}\otimes I^{\otimes^n}\right) CU^{2^{m-1}}\times CU^{2^{m-2}} 
\\
 & \dots CU^{2^1}\times CU^{2^0} \left(U_{QFT}\otimes H^{\otimes^n}\right).
 \end{split}
\end{equation}
\item The iteration operator is defined as: $Q=U_{\psi_3}^\perp U_f$. Apply this operator $O(\frac{1}{|\alpha_x|})$ times to make 
the  success probability for the target state maximum. 
\end{itemize}
\end{tcolorbox}
Note that if there are more than one marked item: i.e. the first $q$ number of eigenvalues are marked, then the success probability at the end of the phase estimation part is the following:
\begin{equation}
\sum_{x=0}^{q-1} |\alpha_x|^2.
\end{equation} 
Therefore, the required  maximum number of runs is bounded by 
\begin{equation}
O\left(\frac{1}{\sqrt{\sum_{x=0}^{q-1} |\alpha_x|^2}}\right).
\end{equation}

\subsection{The Circuit Implementation}
The circuit for the phase estimation part of the algorithm represented by the operator $U_{PEA}$ is the same as the well known circuit for the phase estimation procedure (see the related chapter in Ref.\cite{Kaye2006}). 
In addition to $U_{PEA}$, the implementation of the whole algorithm requires  $U_f$ and $U_0^\perp$:

$U_f$ is a logical operation that marks the eigenvalue if it is greater (or less) than the bounds of the given range. It operates only on the first register and  defined in the following form 
\begin{equation}
\label{Eq:UfGeneral}
U_f=I^{\otimes^m}-2\ket{\bf i}\bra{\bf i},
\end{equation}
where $m$ is the number of qubits and $\ket{\bf i}$ defines which states to be marked. For instance, to mark the states corresponding to the phases greater than 0.5,  we first form $\bra{\bf i}$ in a way that 1 is put to the states which gives a value greater than 0.5 and normalize the formed vector. 
For m=3, 
the obtainable quantum states and the phase values  are:
\ket{000}=1,
\ket{001}=0.125, 
\ket{010}=0.25, 
\ket{011}=0.375, 
\ket{100}=0.5,
\ket{101}=0.625,
\ket{110}=0.750,
\ket{111}=0.875. Therefore, we define 
$\bra{\bf i}=\frac{1}{\sqrt{8}}
\left(\begin{matrix}
 1& 0& 0& 0& 1& 1& 1& 1
\end{matrix}\right)$ and 
use Eq.(\ref{Eq:UfGeneral}) to obtain the reflection operator $U_f$. 
$U_f$ marks only the states corresponding to eigenvalues since the other states does not exist in the superposition: 
 if, for instance, the eigenvalues of the operator greater than 0.5 are  1 and 0.75, then only the states corresponding to these values are marked.
Also note that $U_f$ is  a special case of the general Householder transformation (or reflection). 
Householder transformations are used as ingredients to simulate arbitrary unitary matrices \cite{Bullock2005asymptotically,Ivanov2006engineering,Urias2015householder,Ivanov2008synthesis} and hence  can be efficiently simulatable on quantum computers.

The operator $U_0^\perp$ is also a reflection,  the same as the diffusion operator used in the Grover's algorithm, and can be simulated by using a controlled quantum phase gate.

\section{Numerical Example}

Here, as an illustrative example, we shall use a simple $4$ x $4$ system with the eigenvalues  chosen as:
\begin{equation}
\lambda_0 = 1, \lambda_1 = 0.25, \lambda_2 =  0.5, \text{and\ }  \lambda_3 = 0.75. 
\end{equation}
The columns of the following random orthogonal matrix are assigned as the eigenvectors of the system associated with the above eigenvalues:
\begin{equation}
\left(\begin{matrix}
   -0.6330 &  0.5361&   -0.3094&   -0.4649\\
   -0.4874 &  0.0806 &  -0.1501  &  0.8564\\
    0.0906 &  -0.4553 &  -0.8836 &  -0.0604\\
    0.5946 &   0.7062  & -0.3177   & 0.2163
\end{matrix}\right).
\end{equation}
With the initial equal superposition state, the probabilities to see any of the eigenvalues in the output of the phase estimation algorithm are determined by the squares of the normalized sums of the eigenvector elements: i.e. $|\alpha_1|^2 =0.0473, 
|\alpha_2|^2 = 0.1882,
  |\alpha_3|^2 = 0.6896,$ and
   $|\alpha_4|^2 = 0.0749$.

\begin{figure}
\includegraphics[width=3.5in]{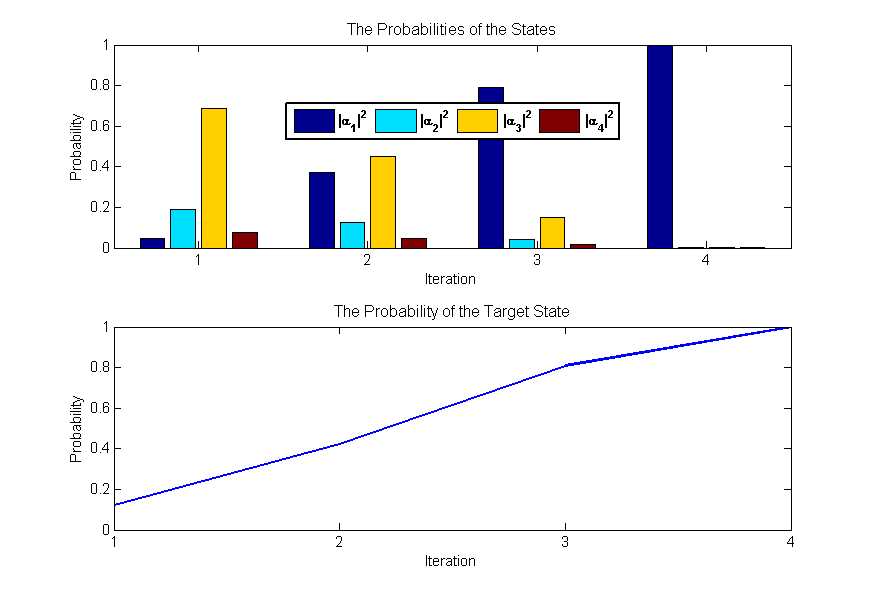}
\caption{The representation of the probability changes in each application of the amplitude amplification to the output of the phase estimation algorithm. The amplitude of the state representing the phase for the eigenvalue 1 is amplified. The x-axis represents the iteration number while the y-axis shows the probabilities of measuring the first register in any of the four states.}
\label{Fig1Principal}
\end{figure}

In order to increase the amplitude of the eigenvalue 1 and the associated eigenvector, we first define $U_f$ as:
\begin{equation}
U_f=I^{\otimes^2}-2\ket{\bf 0}\bra{\bf 0},
\end{equation}
where $\bra{\bf 0}=[1\ 0\ 0\ 0]$. $U_f$ marks the states when the first register is equal to \ket{00} which indicates $\lambda_0$. We then define $U_{\psi}^\perp$:
\begin{equation}
U_{\psi}^\perp
= 
2\ket{\psi}\bra{\psi}-I^{\otimes^{4}} 
=
U_{PEA}U_{0}^\perp U_{PEA}^\dagger,
\end{equation}
The combination of the above two operators, $Q=U_{\psi}^\perp U_f$, is used for the amplitude amplification. After three consecutive applications of $Q$, the probability goes to the maximum as shown in Fig.\ref{Fig1Principal}.

\begin{figure}
\includegraphics[width=3.5in]{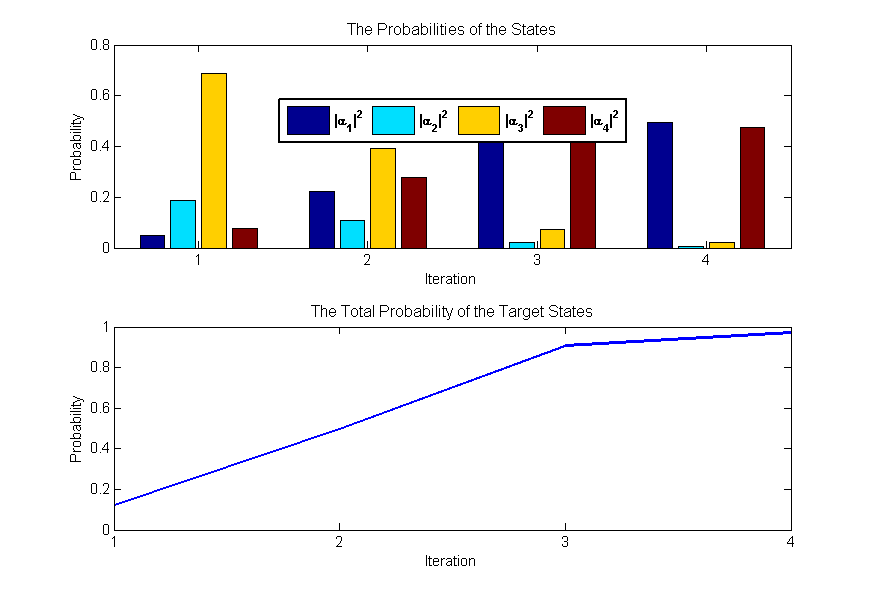}
\caption{The representation of the probability changes in each application of the amplitude amplification to the output of the phase estimation algorithm. The amplitude of the states representing the phases for the eigenvalue 1 and 0.75 is amplified. The x-axis represents the iteration number while the y-axis shows the probabilities of measuring the first register in any of the four states.}
\label{Fig2Principal}
\end{figure}

In the second example, we increase the amplitudes of the eigenvalues 1 and 0.75 by defining $U_f$ as follows:
\begin{equation}
U_f=I^{\otimes^2}-2\ket{\bf i}\bra{\bf i},
\end{equation}
where $\bra{\bf i}= \frac{1}{\sqrt{2}}[1\ 0\ 0\ 1]$. In this case, we mark the amplitudes when the first register is in either \ket{00} or \ket{11} states. The same number of the applications of $Q=U_{\psi}^\perp U_f$ again brings the final state into the superposition of the desired eigenvalues and the eigenvectors: $\approx \beta_1 \ket{00}\ket{\varphi_1} + \beta_4 \ket{11}\ket{\varphi_4}$, where $\beta_1$ and $\beta_2$ are the amplified amplitudes. The probability changes are shown in Fig.\ref{Fig2Principal}.

\section{Discussion and Future Direction}
\subsection{Determining a Range for the Eigenvalues}
The method can be used to study the eigenvectors corresponding to the extremal eigenvalues of an operator by using the upper and lower bounds for these eigenvalues (for the works on the eigenvalue bounds, see e.g. \cite{Higham20035,Diaconis1991,wu2005bounds}). 
It can also be  used to study  the eigenvectors corresponding to a group of eigenvalues for which the upper and the lower limits are known. 
Maximum eigenvalue of a matrix lies between the maximum and minimum column sums due to the correlation of the different matrix norms. 
There are also different bounds for the eigenvalues in specific problems.

One of the methods which can be used to determine bounds for eigenvalues is the Gershgorin theorem \cite{Wilkinson1965algebraic,Stewart1975gershgorin}:
For the each $i$th column ( or row) of a complex matrix, a Gershgorin disc is defined by assigning the diagonal entry on the $i$th column as the center 
and the  absolute sum of the non-diagonal elements thereof as  the radius. 
The Gershgorin theorem indicates that every eigenvalue of the matrix lies within at least one of such discs. And the union of $k$ number of discs disjoint from the rest contains exactly $k$ number of eigenvalues.  As a result, the  eigenvalues lies within the certain range determined from these discs can be used in the method to study the corresponding eigenvectors.

\subsection{Number of Iterations}
\begin{figure*}
\includegraphics[width=5in]{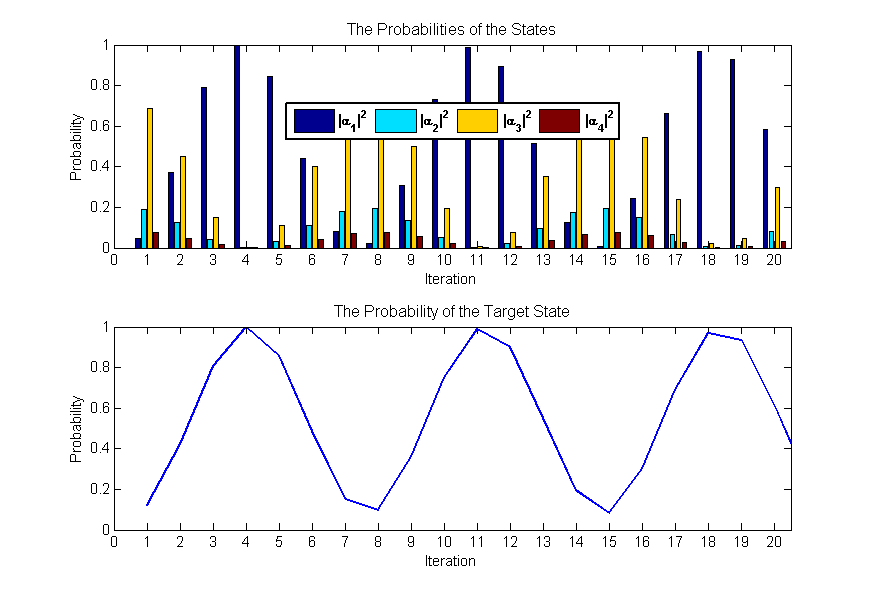}
\caption{The representation of the oscillation of the probability with the successive applications of the amplitude amplification: The amplitude amplification is applied to the output of the phase estimation algorithm. The x-axis represents the iteration number while the y-axis shows the probabilities of measuring the first register in any of the four states.}
\label{FigOscilation}
\end{figure*}

Determining the right number of iterations is one of the most important requirement for the success of the method  since the success probability for the desired states oscillates by the applications of the amplitude amplification. 
This oscillation is shown in Fig.\ref{FigOscilation}, where the probability becomes maximum after four applications of the amplitude amplification but then it starts to oscillate. 
The number of iterations in the amplitude amplification part is determined by $O(\frac{1}{|\alpha_x|})$, where $\alpha_x$ is the amplitude of the desired quantum state. 
 The amplitudes, $\alpha_j$s, at the end of the phase estimation algorithm are the sum of the eigenvector elements. Therefore, without knowing the eigenvector components,  it is impossible to know exact amplitudes. 
 
 However, there exists a method called QSearch (Quantum Search) for searching without knowing the success probabilities \cite{Kaye2006}. In our case, QSearch can be used as follows: 
 After every iteration of the amplitude amplification, the first register holding the superposition of the eigenvalues is measured. If the measurement result is a value in the range $\left[ a, b \right]$ used to construct $U_f$, the superposition of the desired eigenvectors is successfully formed  on the second register. 
 If it is not in the range, then we apply the amplitude amplification again. 
 If there is no desired output obtained in the $\Theta(N)$ number of applications, then we conclude that there is no eigenvalue falling in the range $\left[ a, b \right]$.
  Note that this algorithm uses $O(1/\theta)$ iterations instead of $\frac{\pi}{4\theta}$.
Also note that this  can also be used to estimate an eigenvalue, given that it is the only eigenvalue in the defined range. 

\subsection{Differences to the Classical PCA}
 In classical PCA, one obtains the most significant $q$ number of eigenvalues and do the analysis on  the obtained associated principal eigenvectors by using some weighted or unweighted combination of these eigenvectors either in matrix or in vector forms. 
In some sense, the sum of the $q$ largest eigenvalues are maximized (or the sum of the rest of the eigenvalues are minimized.). Therefore, for different choice of $q$, PCA describes a different maximization or minimization problem \cite{De2005nonlinear}.

However, in the quantum case, instead of $q$ number of principal eigenvectors, we deal with the ones associated with the eigenvalues defined in a range. For instance, if all the eigenvectors are less than one, we can choose the eigenvectors greater than 0.75. 
In some sense, we maximize the sum of the eigenvalues falling into the defined range, which describes a maximization problem similar to but not quite the same as the classical case. Therefore, a different choice of the lower limit for the eigenvalues would describe a different maximization problem. 
Using some eigenvalue bounds and properties, e.g. the Gershgorin theorem, we may also be able to determine the number of eigenvalues falls in the defined range. 

In addition, here, we find the combination of the eigenvectors in the given range rather than $q$ number of largest eigenvalues as in the classical PCA. As explained above, this defines a slightly different maximization problem than the classical PCA. Thus,   the comparison analyses of the applications of the quantum and the classical PCAs to different problems are necessary. The superposition of the eigenstates obtained in the quantum case can also be considered as the weighted linear combination of the eigenvectors in which  the weights are determined by the initial amplitudes $\alpha_j$s. 

\subsection{Application to data}
In this paper, we have assumed the operator $U$ and its power can be efficiently implementable. 
In PCA, the dimension of the data  is generally very high. 
When the Hamiltonian, $\mathcal{H}$, in $U=e^{i\mathcal{H}}$ is sparse, it is known that $U$ can be simulated in polynomial time in the number of qubits \cite{Berry2007sparse, Andrew2011}.  
This is generally not the case for dense matrices. 
Nonetheless,  efficient implementation methods which are specific to some applications may be described. 
For instance, if the dataset can be represented as a sum of simple commuting terms, then the circuit for each term can be found easily. However, this ease of representation is not expected to be exactly attainable in the majority of cases. 
A frequently used method to attain an approximate time-evolution operator of an Hamiltonian  is the Trotter-Suziki decomposition \cite{Trotter1959product,Suzuki1976generalized}. 
In the decomposition, while the required number of gates is at least as many as the number of terms used in the Hamiltonian, the error in the approximation increases  as some power of the number of terms \cite{Poulin2015trotter}.
 It is also known that using additional qubits ease the design of a quantum circuit for a given operator (e.g. see \cite{Daskin2012universal,Daskin2014universal}). 
In the possible application,  a solution may be obtained on the chosen some substate  with the help of additional qubits.

\subsection{Processing the Output State}
Obtaining a quantum state is not a trivial task. To get meaningful information from the quantum state or process the output state on quantum computers, one can benefit from different measurement schemes: e.g. in Ref. \cite{Daskin2015ppi}, we have described a method for the alignment of protein-protein interaction networks and showed how to extract the solution of the alignment efficiently from the measurement statistics: 
In that case,  the largest magnitudes of the amplitudes are sufficient to predict an optimal solution for the alignment. 
Therefore, some similar schemes can be used for the other problems 
where extracting the optimal solution from the output state does not require the topology of the whole output state and 
 the significant magnitudes can be used to predict a solution.
 In some cases, the qubits can  also be grouped to use different conditional measurement schemes to obtain meaningful information for the solution.
  
  However, It is known that obtaining the topology of a whole quantum state requires exponential number of computational steps. 
  For the problems whose solutions necessitate the whole vector coefficients,  there is a need for research in designing application-specific algorithms to process the obtained quantum eigenstate mostly on the quantum computers.   

\subsection{Possible Applications}
We believe the method described here can be applied to quantum images and may pave the ways to devise algorithms for the applications where principal component analysis is used successfully: e.g. pattern and face recognition.

In Ref.\cite{Daskin2012universal,Daskin2014universal}, a general circuit design consisting of $2log_2N$ qubits and $O(N^{2})$ quantum gates is given for the simulation of a matrix of order $N$ not necessarily Hermitian. In particular, it is shown that any matrix can be simulated on certain predetermined states by using $log_2N$ ancilla qubits. However, although the circuit design method requires only a linear number of computational steps, $O(N)$, the success probability decreases exponentially with the system size, i.e. $\frac{1}{\sqrt{N}}$.  Since the states in the simulation are predetermined, the amplitude amplification method described here can also be applied $O(\sqrt{N})$ times to the circuit design given in Ref.\cite{Daskin2014universal} so as to increase the success probability upto 1. This modification would also increase the complexity, the number of quantum gates, of the circuit to $O(\sqrt{N}N^{2})$. In addition, a similar modified circuit design may be employed in the principal component analysis to simulate the data matrix. 

\section{Conclusion}
In this paper, we have shown a method by employing the quantum phase estimation algorithm together with the amplitude amplification to obtain the superposition of the principal eigenvectors of a matrix. 
We have also discussed how to study the eigenvalues defined in a certain range and their corresponding eigenvectors by using this method. Finally, we have shown a numerical example and discussed possible applications. The method paves the way to do principal component analysis on quantum computers.

\section{Acknowledgment}
This work is supported by TUBITAK under the project number 115E747.

\bibliography{reference}

\begin{thebibliography}{43}%
\makeatletter
\providecommand \@ifxundefined [1]{%
 \@ifx{#1\undefined}
}%
\providecommand \@ifnum [1]{%
 \ifnum #1\expandafter \@firstoftwo
 \else \expandafter \@secondoftwo
 \fi
}%
\providecommand \@ifx [1]{%
 \ifx #1\expandafter \@firstoftwo
 \else \expandafter \@secondoftwo
 \fi
}%
\providecommand \natexlab [1]{#1}%
\providecommand \enquote  [1]{``#1''}%
\providecommand \bibnamefont  [1]{#1}%
\providecommand \bibfnamefont [1]{#1}%
\providecommand \citenamefont [1]{#1}%
\providecommand \href@noop [0]{\@secondoftwo}%
\providecommand \href [0]{\begingroup \@sanitize@url \@href}%
\providecommand \@href[1]{\@@startlink{#1}\@@href}%
\providecommand \@@href[1]{\endgroup#1\@@endlink}%
\providecommand \@sanitize@url [0]{\catcode `\\12\catcode `\$12\catcode
  `\&12\catcode `\#12\catcode `\^12\catcode `\_12\catcode `\%12\relax}%
\providecommand \@@startlink[1]{}%
\providecommand \@@endlink[0]{}%
\providecommand \url  [0]{\begingroup\@sanitize@url \@url }%
\providecommand \@url [1]{\endgroup\@href {#1}{\urlprefix }}%
\providecommand \urlprefix  [0]{URL }%
\providecommand \Eprint [0]{\href }%
\providecommand \doibase [0]{http://dx.doi.org/}%
\providecommand \selectlanguage [0]{\@gobble}%
\providecommand \bibinfo  [0]{\@secondoftwo}%
\providecommand \bibfield  [0]{\@secondoftwo}%
\providecommand \translation [1]{[#1]}%
\providecommand \BibitemOpen [0]{}%
\providecommand \bibitemStop [0]{}%
\providecommand \bibitemNoStop [0]{.\EOS\space}%
\providecommand \EOS [0]{\spacefactor3000\relax}%
\providecommand \BibitemShut  [1]{\csname bibitem#1\endcsname}%
\let\auto@bib@innerbib\@empty
\bibitem [{\citenamefont {F.R.S.}(1901)}]{Pearson1901}%
  \BibitemOpen
  \bibfield  {author} {\bibinfo {author} {\bibfnamefont {Karl~Pearson}\
  \bibnamefont {F.R.S.}},\ }\bibfield  {title} {\enquote {\bibinfo {title}
  {Liii. on lines and planes of closest fit to systems of points in space},}\
  }\href {\doibase 10.1080/14786440109462720} {\bibfield  {journal} {\bibinfo
  {journal} {Philosophical Magazine Series 6}\ }\textbf {\bibinfo {volume}
  {2}},\ \bibinfo {pages} {559--572} (\bibinfo {year} {1901})},\ \Eprint
  {http://arxiv.org/abs/http://dx.doi.org/10.1080/14786440109462720}
  {http://dx.doi.org/10.1080/14786440109462720} \BibitemShut {NoStop}%
\bibitem [{\citenamefont {Jolliffe}(2002)}]{Jolliffe2002}%
  \BibitemOpen
  \bibfield  {author} {\bibinfo {author} {\bibfnamefont {Ian}\ \bibnamefont
  {Jolliffe}},\ }\href@noop {} {\emph {\bibinfo {title} {Principal component
  analysis}}}\ (\bibinfo  {publisher} {Wiley Online Library},\ \bibinfo {year}
  {2002})\BibitemShut {NoStop}%
\bibitem [{\citenamefont {Tipping}\ and\ \citenamefont
  {Bishop}(1999)}]{Tipping1999probabilistic}%
  \BibitemOpen
  \bibfield  {author} {\bibinfo {author} {\bibfnamefont {Michael~E}\
  \bibnamefont {Tipping}}\ and\ \bibinfo {author} {\bibfnamefont
  {Christopher~M}\ \bibnamefont {Bishop}},\ }\bibfield  {title} {\enquote
  {\bibinfo {title} {Probabilistic principal component analysis},}\ }\href@noop
  {} {\bibfield  {journal} {\bibinfo  {journal} {Journal of the Royal
  Statistical Society: Series B (Statistical Methodology)}\ }\textbf {\bibinfo
  {volume} {61}},\ \bibinfo {pages} {611--622} (\bibinfo {year}
  {1999})}\BibitemShut {NoStop}%
\bibitem [{\citenamefont {de~Leeuw}(2005)}]{De2005nonlinear}%
  \BibitemOpen
  \bibfield  {author} {\bibinfo {author} {\bibfnamefont {Jan}\ \bibnamefont
  {de~Leeuw}},\ }\bibfield  {title} {\enquote {\bibinfo {title} {Nonlinear
  principal component analysis},}\ }\href@noop {} {\bibfield  {journal}
  {\bibinfo  {journal} {Department of Statistics, UCLA}\ } (\bibinfo {year}
  {2005})}\BibitemShut {NoStop}%
\bibitem [{\citenamefont {Sch{\"o}lkopf}\ \emph {et~al.}(1997)\citenamefont
  {Sch{\"o}lkopf}, \citenamefont {Smola},\ and\ \citenamefont
  {M{\"u}ller}}]{Scholkopf1997kernel}%
  \BibitemOpen
  \bibfield  {author} {\bibinfo {author} {\bibfnamefont {Bernhard}\
  \bibnamefont {Sch{\"o}lkopf}}, \bibinfo {author} {\bibfnamefont {Alexander}\
  \bibnamefont {Smola}}, \ and\ \bibinfo {author} {\bibfnamefont
  {Klaus-Robert}\ \bibnamefont {M{\"u}ller}},\ }\bibfield  {title} {\enquote
  {\bibinfo {title} {Kernel principal component analysis},}\ }in\ \href@noop {}
  {\emph {\bibinfo {booktitle} {Artificial Neural Networks—ICANN'97}}}\
  (\bibinfo  {publisher} {Springer},\ \bibinfo {year} {1997})\ pp.\ \bibinfo
  {pages} {583--588}\BibitemShut {NoStop}%
\bibitem [{\citenamefont {Kambhatla}\ and\ \citenamefont
  {Leen}(1997)}]{Kambhatla1997dimension}%
  \BibitemOpen
  \bibfield  {author} {\bibinfo {author} {\bibfnamefont {Nandakishore}\
  \bibnamefont {Kambhatla}}\ and\ \bibinfo {author} {\bibfnamefont {Todd~K}\
  \bibnamefont {Leen}},\ }\bibfield  {title} {\enquote {\bibinfo {title}
  {Dimension reduction by local principal component analysis},}\ }\href@noop {}
  {\bibfield  {journal} {\bibinfo  {journal} {Neural Computation}\ }\textbf
  {\bibinfo {volume} {9}},\ \bibinfo {pages} {1493--1516} (\bibinfo {year}
  {1997})}\BibitemShut {NoStop}%
\bibitem [{\citenamefont {Vidal}\ \emph {et~al.}(2005)\citenamefont {Vidal},
  \citenamefont {Ma},\ and\ \citenamefont {Sastry}}]{Vidal2005generalized}%
  \BibitemOpen
  \bibfield  {author} {\bibinfo {author} {\bibfnamefont {Rene}\ \bibnamefont
  {Vidal}}, \bibinfo {author} {\bibfnamefont {Yi}~\bibnamefont {Ma}}, \ and\
  \bibinfo {author} {\bibfnamefont {Shankar}\ \bibnamefont {Sastry}},\
  }\bibfield  {title} {\enquote {\bibinfo {title} {Generalized principal
  component analysis (gpca)},}\ }\href@noop {} {\bibfield  {journal} {\bibinfo
  {journal} {Pattern Analysis and Machine Intelligence, IEEE Transactions on}\
  }\textbf {\bibinfo {volume} {27}},\ \bibinfo {pages} {1945--1959} (\bibinfo
  {year} {2005})}\BibitemShut {NoStop}%
\bibitem [{\citenamefont {Bishop}(2006)}]{Bishop2006pattern}%
  \BibitemOpen
  \bibfield  {author} {\bibinfo {author} {\bibfnamefont {Christopher~M}\
  \bibnamefont {Bishop}},\ }\href@noop {} {\emph {\bibinfo {title} {Pattern
  recognition and machine learning}}}\ (\bibinfo  {publisher} {springer},\
  \bibinfo {year} {2006})\BibitemShut {NoStop}%
\bibitem [{\citenamefont {Shlens}(2014)}]{Shlens2014tutorial}%
  \BibitemOpen
  \bibfield  {author} {\bibinfo {author} {\bibfnamefont {Jonathon}\
  \bibnamefont {Shlens}},\ }\bibfield  {title} {\enquote {\bibinfo {title} {A
  tutorial on principal component analysis},}\ }\href@noop {} {\bibfield
  {journal} {\bibinfo  {journal} {arXiv preprint arXiv:1404.1100}\ } (\bibinfo
  {year} {2014})}\BibitemShut {NoStop}%
\bibitem [{\citenamefont {Kitaev}(1996)}]{Kitaev1996}%
  \BibitemOpen
  \bibfield  {author} {\bibinfo {author} {\bibfnamefont {Alexei}\ \bibnamefont
  {Kitaev}},\ }\bibfield  {title} {\enquote {\bibinfo {title} {{Quantum
  measurements and the Abelian Stabilizer Problem}},}\ }\href@noop {}
  {\bibfield  {journal} {\bibinfo  {journal} {Electronic Colloquium on
  Computational Complexity (ECCC)}\ }\textbf {\bibinfo {volume} {3}} (\bibinfo
  {year} {1996})}\BibitemShut {NoStop}%
\bibitem [{\citenamefont {Papageorgiou}\ \emph {et~al.}(2013)\citenamefont
  {Papageorgiou}, \citenamefont {Petras}, \citenamefont {Traub},\ and\
  \citenamefont {Zhang}}]{Papageorgiou2013fast}%
  \BibitemOpen
  \bibfield  {author} {\bibinfo {author} {\bibfnamefont {Anargyros}\
  \bibnamefont {Papageorgiou}}, \bibinfo {author} {\bibfnamefont {Iasonas}\
  \bibnamefont {Petras}}, \bibinfo {author} {\bibfnamefont {J}~\bibnamefont
  {Traub}}, \ and\ \bibinfo {author} {\bibfnamefont {Chi}\ \bibnamefont
  {Zhang}},\ }\bibfield  {title} {\enquote {\bibinfo {title} {A fast algorithm
  for approximating the ground state energy on a quantum computer},}\
  }\href@noop {} {\bibfield  {journal} {\bibinfo  {journal} {Mathematics of
  Computation}\ }\textbf {\bibinfo {volume} {82}},\ \bibinfo {pages}
  {2293--2304} (\bibinfo {year} {2013})}\BibitemShut {NoStop}%
\bibitem [{\citenamefont {Daskin}\ \emph
  {et~al.}(2014{\natexlab{a}})\citenamefont {Daskin}, \citenamefont {Grama},\
  and\ \citenamefont {Kais}}]{Daskin2014mna}%
  \BibitemOpen
  \bibfield  {author} {\bibinfo {author} {\bibfnamefont {Anmer}\ \bibnamefont
  {Daskin}}, \bibinfo {author} {\bibfnamefont {Ananth}\ \bibnamefont {Grama}},
  \ and\ \bibinfo {author} {\bibfnamefont {Sabre}\ \bibnamefont {Kais}},\
  }\bibfield  {title} {\enquote {\bibinfo {title} {Multiple network alignment
  on quantum computers},}\ }\href {\doibase 10.1007/s11128-014-0818-7}
  {\bibfield  {journal} {\bibinfo  {journal} {Quantum Information Processing}\
  }\textbf {\bibinfo {volume} {13}},\ \bibinfo {pages} {2653--2666} (\bibinfo
  {year} {2014}{\natexlab{a}})}\BibitemShut {NoStop}%
\bibitem [{\citenamefont {Daskin}(2015)}]{Daskin2015ppi}%
  \BibitemOpen
  \bibfield  {author} {\bibinfo {author} {\bibfnamefont {Anmer}\ \bibnamefont
  {Daskin}},\ }\bibfield  {title} {\enquote {\bibinfo {title} {Quantum isorank:
  Efficient alignment of multiple ppi networks},}\ }\href@noop {} {\bibfield
  {journal} {\bibinfo  {journal} {arXiv preprint arXiv:1506.05905}\ } (\bibinfo
  {year} {2015})}\BibitemShut {NoStop}%
\bibitem [{\citenamefont {Harrow}\ \emph {et~al.}(2009)\citenamefont {Harrow},
  \citenamefont {Hassidim},\ and\ \citenamefont {Lloyd}}]{Harrow2009quantum}%
  \BibitemOpen
  \bibfield  {author} {\bibinfo {author} {\bibfnamefont {Aram~W}\ \bibnamefont
  {Harrow}}, \bibinfo {author} {\bibfnamefont {Avinatan}\ \bibnamefont
  {Hassidim}}, \ and\ \bibinfo {author} {\bibfnamefont {Seth}\ \bibnamefont
  {Lloyd}},\ }\bibfield  {title} {\enquote {\bibinfo {title} {Quantum algorithm
  for linear systems of equations},}\ }\href@noop {} {\bibfield  {journal}
  {\bibinfo  {journal} {Physical review letters}\ }\textbf {\bibinfo {volume}
  {103}},\ \bibinfo {pages} {150502} (\bibinfo {year} {2009})}\BibitemShut
  {NoStop}%
\bibitem [{\citenamefont {Rebentrost}\ \emph {et~al.}(2014)\citenamefont
  {Rebentrost}, \citenamefont {Mohseni},\ and\ \citenamefont
  {Lloyd}}]{Rebentrost2014quantum}%
  \BibitemOpen
  \bibfield  {author} {\bibinfo {author} {\bibfnamefont {Patrick}\ \bibnamefont
  {Rebentrost}}, \bibinfo {author} {\bibfnamefont {Masoud}\ \bibnamefont
  {Mohseni}}, \ and\ \bibinfo {author} {\bibfnamefont {Seth}\ \bibnamefont
  {Lloyd}},\ }\bibfield  {title} {\enquote {\bibinfo {title} {Quantum support
  vector machine for big data classification},}\ }\href@noop {} {\bibfield
  {journal} {\bibinfo  {journal} {Physical review letters}\ }\textbf {\bibinfo
  {volume} {113}},\ \bibinfo {pages} {130503} (\bibinfo {year}
  {2014})}\BibitemShut {NoStop}%
\bibitem [{\citenamefont {Wittek}(2014)}]{Wittek2014quantum}%
  \BibitemOpen
  \bibfield  {author} {\bibinfo {author} {\bibfnamefont {Peter}\ \bibnamefont
  {Wittek}},\ }\href@noop {} {\emph {\bibinfo {title} {Quantum Machine
  Learning: What Quantum Computing Means to Data Mining}}}\ (\bibinfo
  {publisher} {Academic Press},\ \bibinfo {year} {2014})\BibitemShut {NoStop}%
\bibitem [{\citenamefont {Schuld}\ \emph {et~al.}(2014)\citenamefont {Schuld},
  \citenamefont {Sinayskiy},\ and\ \citenamefont
  {Petruccione}}]{Schuld2014quantum}%
  \BibitemOpen
  \bibfield  {author} {\bibinfo {author} {\bibfnamefont {Maria}\ \bibnamefont
  {Schuld}}, \bibinfo {author} {\bibfnamefont {Ilya}\ \bibnamefont
  {Sinayskiy}}, \ and\ \bibinfo {author} {\bibfnamefont {Francesco}\
  \bibnamefont {Petruccione}},\ }\bibfield  {title} {\enquote {\bibinfo {title}
  {Quantum computing for pattern classification},}\ }in\ \href@noop {} {\emph
  {\bibinfo {booktitle} {PRICAI 2014: Trends in Artificial Intelligence}}}\
  (\bibinfo  {publisher} {Springer},\ \bibinfo {year} {2014})\ pp.\ \bibinfo
  {pages} {208--220}\BibitemShut {NoStop}%
\bibitem [{\citenamefont {Neven}\ \emph {et~al.}(2008)\citenamefont {Neven},
  \citenamefont {Denchev}, \citenamefont {Rose},\ and\ \citenamefont
  {Macready}}]{Neven2008training}%
  \BibitemOpen
  \bibfield  {author} {\bibinfo {author} {\bibfnamefont {Hartmut}\ \bibnamefont
  {Neven}}, \bibinfo {author} {\bibfnamefont {Vasil~S}\ \bibnamefont
  {Denchev}}, \bibinfo {author} {\bibfnamefont {Geordie}\ \bibnamefont {Rose}},
  \ and\ \bibinfo {author} {\bibfnamefont {William~G}\ \bibnamefont
  {Macready}},\ }\bibfield  {title} {\enquote {\bibinfo {title} {Training a
  binary classifier with the quantum adiabatic algorithm},}\ }\href@noop {}
  {\bibfield  {journal} {\bibinfo  {journal} {arXiv preprint arXiv:0811.0416}\
  } (\bibinfo {year} {2008})}\BibitemShut {NoStop}%
\bibitem [{\citenamefont {Neven}\ \emph {et~al.}(2009)\citenamefont {Neven},
  \citenamefont {Denchev}, \citenamefont {Rose},\ and\ \citenamefont
  {Macready}}]{Neven2009training}%
  \BibitemOpen
  \bibfield  {author} {\bibinfo {author} {\bibfnamefont {Hartmut}\ \bibnamefont
  {Neven}}, \bibinfo {author} {\bibfnamefont {Vasil~S}\ \bibnamefont
  {Denchev}}, \bibinfo {author} {\bibfnamefont {Geordie}\ \bibnamefont {Rose}},
  \ and\ \bibinfo {author} {\bibfnamefont {William~G}\ \bibnamefont
  {Macready}},\ }\bibfield  {title} {\enquote {\bibinfo {title} {Training a
  large scale classifier with the quantum adiabatic algorithm},}\ }\href@noop
  {} {\bibfield  {journal} {\bibinfo  {journal} {arXiv preprint
  arXiv:0912.0779}\ } (\bibinfo {year} {2009})}\BibitemShut {NoStop}%
\bibitem [{\citenamefont {A{\"\i}meur}\ \emph {et~al.}(2013)\citenamefont
  {A{\"\i}meur}, \citenamefont {Brassard},\ and\ \citenamefont
  {Gambs}}]{Aimeur2013quantum}%
  \BibitemOpen
  \bibfield  {author} {\bibinfo {author} {\bibfnamefont {Esma}\ \bibnamefont
  {A{\"\i}meur}}, \bibinfo {author} {\bibfnamefont {Gilles}\ \bibnamefont
  {Brassard}}, \ and\ \bibinfo {author} {\bibfnamefont {S{\'e}bastien}\
  \bibnamefont {Gambs}},\ }\bibfield  {title} {\enquote {\bibinfo {title}
  {Quantum speed-up for unsupervised learning},}\ }\href@noop {} {\bibfield
  {journal} {\bibinfo  {journal} {Machine Learning}\ }\textbf {\bibinfo
  {volume} {90}},\ \bibinfo {pages} {261--287} (\bibinfo {year}
  {2013})}\BibitemShut {NoStop}%
\bibitem [{\citenamefont {Lloyd}\ \emph
  {et~al.}(2014{\natexlab{a}})\citenamefont {Lloyd}, \citenamefont
  {Garnerone},\ and\ \citenamefont {Zanardi}}]{Lloyd2014topological}%
  \BibitemOpen
  \bibfield  {author} {\bibinfo {author} {\bibfnamefont {Seth}\ \bibnamefont
  {Lloyd}}, \bibinfo {author} {\bibfnamefont {Silvano}\ \bibnamefont
  {Garnerone}}, \ and\ \bibinfo {author} {\bibfnamefont {Paolo}\ \bibnamefont
  {Zanardi}},\ }\bibfield  {title} {\enquote {\bibinfo {title} {Quantum
  algorithms for topological and geometric analysis of big data},}\ }\href@noop
  {} {\bibfield  {journal} {\bibinfo  {journal} {arXiv preprint
  arXiv:1408.3106}\ } (\bibinfo {year} {2014}{\natexlab{a}})}\BibitemShut
  {NoStop}%
\bibitem [{\citenamefont {Lloyd}\ \emph
  {et~al.}(2014{\natexlab{b}})\citenamefont {Lloyd}, \citenamefont {Mohseni},\
  and\ \citenamefont {Rebentrost}}]{lloyd2014qQPCA}%
  \BibitemOpen
  \bibfield  {author} {\bibinfo {author} {\bibfnamefont {Seth}\ \bibnamefont
  {Lloyd}}, \bibinfo {author} {\bibfnamefont {Masoud}\ \bibnamefont {Mohseni}},
  \ and\ \bibinfo {author} {\bibfnamefont {Patrick}\ \bibnamefont
  {Rebentrost}},\ }\bibfield  {title} {\enquote {\bibinfo {title} {Quantum
  principal component analysis},}\ }\href@noop {} {\bibfield  {journal}
  {\bibinfo  {journal} {Nature Physics}\ }\textbf {\bibinfo {volume} {10}},\
  \bibinfo {pages} {631--633} (\bibinfo {year}
  {2014}{\natexlab{b}})}\BibitemShut {NoStop}%
\bibitem [{\citenamefont {Grover}(1998)}]{Grover1998}%
  \BibitemOpen
  \bibfield  {author} {\bibinfo {author} {\bibfnamefont {Lov~K}\ \bibnamefont
  {Grover}},\ }\bibfield  {title} {\enquote {\bibinfo {title} {Quantum
  computers can search rapidly by using almost any transformation},}\
  }\href@noop {} {\bibfield  {journal} {\bibinfo  {journal} {Physical Review
  Letters}\ }\textbf {\bibinfo {volume} {80}},\ \bibinfo {pages} {4329}
  (\bibinfo {year} {1998})}\BibitemShut {NoStop}%
\bibitem [{\citenamefont {Mosca}\ \emph {et~al.}(1998)\citenamefont {Mosca}
  \emph {et~al.}}]{Mosca1998quantum}%
  \BibitemOpen
  \bibfield  {author} {\bibinfo {author} {\bibfnamefont {Michele}\ \bibnamefont
  {Mosca}} \emph {et~al.},\ }\bibfield  {title} {\enquote {\bibinfo {title}
  {Quantum searching, counting and amplitude amplification by eigenvector
  analysis},}\ }in\ \href@noop {} {\emph {\bibinfo {booktitle} {MFCS’98
  workshop on Randomized Algorithms}}}\ (\bibinfo {year} {1998})\ pp.\ \bibinfo
  {pages} {90--100}\BibitemShut {NoStop}%
\bibitem [{\citenamefont {Brassard}\ \emph {et~al.}(2002)\citenamefont
  {Brassard}, \citenamefont {Hoyer}, \citenamefont {Mosca},\ and\ \citenamefont
  {Tapp}}]{Brassard2002}%
  \BibitemOpen
  \bibfield  {author} {\bibinfo {author} {\bibfnamefont {Gilles}\ \bibnamefont
  {Brassard}}, \bibinfo {author} {\bibfnamefont {Peter}\ \bibnamefont {Hoyer}},
  \bibinfo {author} {\bibfnamefont {Michele}\ \bibnamefont {Mosca}}, \ and\
  \bibinfo {author} {\bibfnamefont {Alain}\ \bibnamefont {Tapp}},\ }\bibfield
  {title} {\enquote {\bibinfo {title} {Quantum amplitude amplification and
  estimation},}\ }\href@noop {} {\bibfield  {journal} {\bibinfo  {journal}
  {Contemporary Mathematics}\ }\textbf {\bibinfo {volume} {305}},\ \bibinfo
  {pages} {53--74} (\bibinfo {year} {2002})}\BibitemShut {NoStop}%
\bibitem [{\citenamefont {Kaye}\ \emph {et~al.}(2006)\citenamefont {Kaye},
  \citenamefont {Mosca},\ and\ \citenamefont {Laflamme}}]{Kaye2006}%
  \BibitemOpen
  \bibfield  {author} {\bibinfo {author} {\bibfnamefont {Phillip}\ \bibnamefont
  {Kaye}}, \bibinfo {author} {\bibfnamefont {Michele}\ \bibnamefont {Mosca}}, \
  and\ \bibinfo {author} {\bibfnamefont {Raymond}\ \bibnamefont {Laflamme}},\
  }\href@noop {} {\emph {\bibinfo {title} {An introduction to quantum
  computing}}}\ (\bibinfo  {publisher} {Oxford Univ. Press, UK},\ \bibinfo
  {year} {2006})\BibitemShut {NoStop}%
\bibitem [{\citenamefont {Mosca}(2001)}]{Mosca2001139}%
  \BibitemOpen
  \bibfield  {author} {\bibinfo {author} {\bibfnamefont {Michele}\ \bibnamefont
  {Mosca}},\ }\bibfield  {title} {\enquote {\bibinfo {title} {Counting by
  quantum eigenvalue estimation},}\ }\href {\doibase
  http://dx.doi.org/10.1016/S0304-3975(00)00217-6} {\bibfield  {journal}
  {\bibinfo  {journal} {Theoretical Computer Science}\ }\textbf {\bibinfo
  {volume} {264}},\ \bibinfo {pages} {139 -- 153} (\bibinfo {year} {2001})},\
  \bibinfo {note} {mFCS'98}\BibitemShut {NoStop}%
\bibitem [{\citenamefont {Bullock}\ \emph {et~al.}(2005)\citenamefont
  {Bullock}, \citenamefont {O’Leary},\ and\ \citenamefont
  {Brennen}}]{Bullock2005asymptotically}%
  \BibitemOpen
  \bibfield  {author} {\bibinfo {author} {\bibfnamefont {Stephen~S}\
  \bibnamefont {Bullock}}, \bibinfo {author} {\bibfnamefont {Dianne~P}\
  \bibnamefont {O’Leary}}, \ and\ \bibinfo {author} {\bibfnamefont {Gavin~K}\
  \bibnamefont {Brennen}},\ }\bibfield  {title} {\enquote {\bibinfo {title}
  {Asymptotically optimal quantum circuits for d-level systems},}\ }\href@noop
  {} {\bibfield  {journal} {\bibinfo  {journal} {Physical review letters}\
  }\textbf {\bibinfo {volume} {94}},\ \bibinfo {pages} {230502} (\bibinfo
  {year} {2005})}\BibitemShut {NoStop}%
\bibitem [{\citenamefont {Ivanov}\ \emph {et~al.}(2006)\citenamefont {Ivanov},
  \citenamefont {Kyoseva},\ and\ \citenamefont
  {Vitanov}}]{Ivanov2006engineering}%
  \BibitemOpen
  \bibfield  {author} {\bibinfo {author} {\bibfnamefont {Peter~A}\ \bibnamefont
  {Ivanov}}, \bibinfo {author} {\bibfnamefont {ES}~\bibnamefont {Kyoseva}}, \
  and\ \bibinfo {author} {\bibfnamefont {NV}~\bibnamefont {Vitanov}},\
  }\bibfield  {title} {\enquote {\bibinfo {title} {Engineering of arbitrary u
  (n) transformations by quantum householder reflections},}\ }\href@noop {}
  {\bibfield  {journal} {\bibinfo  {journal} {Physical Review A}\ }\textbf
  {\bibinfo {volume} {74}},\ \bibinfo {pages} {022323} (\bibinfo {year}
  {2006})}\BibitemShut {NoStop}%
\bibitem [{\citenamefont {Ur{\'\i}as}\ and\ \citenamefont
  {Qui{\~n}ones}(2015)}]{Urias2015householder}%
  \BibitemOpen
  \bibfield  {author} {\bibinfo {author} {\bibfnamefont {Jes{\'u}s}\
  \bibnamefont {Ur{\'\i}as}}\ and\ \bibinfo {author} {\bibfnamefont {Diego~A}\
  \bibnamefont {Qui{\~n}ones}},\ }\bibfield  {title} {\enquote {\bibinfo
  {title} {Householder methods for quantum circuit design},}\ }\href@noop {}
  {\bibfield  {journal} {\bibinfo  {journal} {Canadian Journal of Physics}\
  }\textbf {\bibinfo {volume} {93}},\ \bibinfo {pages} {1--8} (\bibinfo {year}
  {2015})}\BibitemShut {NoStop}%
\bibitem [{\citenamefont {Ivanov}\ and\ \citenamefont
  {Vitanov}(2008)}]{Ivanov2008synthesis}%
  \BibitemOpen
  \bibfield  {author} {\bibinfo {author} {\bibfnamefont {Peter~A}\ \bibnamefont
  {Ivanov}}\ and\ \bibinfo {author} {\bibfnamefont {Nikolay~V}\ \bibnamefont
  {Vitanov}},\ }\bibfield  {title} {\enquote {\bibinfo {title} {Synthesis of
  arbitrary unitary transformations of collective states of trapped ions by
  quantum householder reflections},}\ }\href@noop {} {\bibfield  {journal}
  {\bibinfo  {journal} {Physical Review A}\ }\textbf {\bibinfo {volume} {77}},\
  \bibinfo {pages} {012335} (\bibinfo {year} {2008})}\BibitemShut {NoStop}%
\bibitem [{\citenamefont {Higham}\ and\ \citenamefont
  {Tisseur}(2003)}]{Higham20035}%
  \BibitemOpen
  \bibfield  {author} {\bibinfo {author} {\bibfnamefont {Nicholas~J.}\
  \bibnamefont {Higham}}\ and\ \bibinfo {author} {\bibfnamefont {Françoise}\
  \bibnamefont {Tisseur}},\ }\bibfield  {title} {\enquote {\bibinfo {title}
  {Bounds for eigenvalues of matrix polynomials},}\ }\href {\doibase
  http://dx.doi.org/10.1016/S0024-3795(01)00316-0} {\bibfield  {journal}
  {\bibinfo  {journal} {Linear Algebra and its Applications}\ }\textbf
  {\bibinfo {volume} {358}},\ \bibinfo {pages} {5 -- 22} (\bibinfo {year}
  {2003})}\BibitemShut {NoStop}%
\bibitem [{\citenamefont {Persi~Diaconis}(1991)}]{Diaconis1991}%
  \BibitemOpen
  \bibfield  {author} {\bibinfo {author} {\bibfnamefont {Daniel~Stroock}\
  \bibnamefont {Persi~Diaconis}},\ }\bibfield  {title} {\enquote {\bibinfo
  {title} {Geometric bounds for eigenvalues of markov chains},}\ }\href
  {http://www.jstor.org/stable/2959624} {\bibfield  {journal} {\bibinfo
  {journal} {The Annals of Applied Probability}\ }\textbf {\bibinfo {volume}
  {1}},\ \bibinfo {pages} {36--61} (\bibinfo {year} {1991})}\BibitemShut
  {NoStop}%
\bibitem [{\citenamefont {Wu}(2005)}]{wu2005bounds}%
  \BibitemOpen
  \bibfield  {author} {\bibinfo {author} {\bibfnamefont {Chai~Wah}\
  \bibnamefont {Wu}},\ }\bibfield  {title} {\enquote {\bibinfo {title} {On
  bounds of extremal eigenvalues of irreducible and m-reducible matrices},}\
  }\href@noop {} {\bibfield  {journal} {\bibinfo  {journal} {Linear algebra and
  its applications}\ }\textbf {\bibinfo {volume} {402}},\ \bibinfo {pages}
  {29--45} (\bibinfo {year} {2005})}\BibitemShut {NoStop}%
\bibitem [{\citenamefont {Wilkinson}(1965)}]{Wilkinson1965algebraic}%
  \BibitemOpen
  \bibfield  {author} {\bibinfo {author} {\bibfnamefont {James~Hardy}\
  \bibnamefont {Wilkinson}},\ }\href@noop {} {\emph {\bibinfo {title} {The
  algebraic eigenvalue problem}}},\ Vol.~\bibinfo {volume} {87}\ (\bibinfo
  {publisher} {Clarendon Press Oxford},\ \bibinfo {year} {1965})\BibitemShut
  {NoStop}%
\bibitem [{\citenamefont {Stewart}(1975)}]{Stewart1975gershgorin}%
  \BibitemOpen
  \bibfield  {author} {\bibinfo {author} {\bibfnamefont {Gilbert~W}\
  \bibnamefont {Stewart}},\ }\bibfield  {title} {\enquote {\bibinfo {title}
  {Gershgorin theory for the generalized eigenvalue problem 𝐴𝑥=
  𝜆𝐵𝑥},}\ }\href@noop {} {\bibfield  {journal} {\bibinfo  {journal}
  {Mathematics of Computation}\ }\textbf {\bibinfo {volume} {29}},\ \bibinfo
  {pages} {600--606} (\bibinfo {year} {1975})}\BibitemShut {NoStop}%
\bibitem [{\citenamefont {Berry}\ \emph {et~al.}(2007)\citenamefont {Berry},
  \citenamefont {Ahokas}, \citenamefont {Cleve},\ and\ \citenamefont
  {Sanders}}]{Berry2007sparse}%
  \BibitemOpen
  \bibfield  {author} {\bibinfo {author} {\bibfnamefont {DominicW.}\
  \bibnamefont {Berry}}, \bibinfo {author} {\bibfnamefont {Graeme}\
  \bibnamefont {Ahokas}}, \bibinfo {author} {\bibfnamefont {Richard}\
  \bibnamefont {Cleve}}, \ and\ \bibinfo {author} {\bibfnamefont {BarryC.}\
  \bibnamefont {Sanders}},\ }\bibfield  {title} {\enquote {\bibinfo {title}
  {Efficient quantum algorithms for simulating sparse hamiltonians},}\ }\href
  {\doibase 10.1007/s00220-006-0150-x} {\bibfield  {journal} {\bibinfo
  {journal} {Communications in Mathematical Physics}\ }\textbf {\bibinfo
  {volume} {270}},\ \bibinfo {pages} {359--371} (\bibinfo {year}
  {2007})}\BibitemShut {NoStop}%
\bibitem [{\citenamefont {Childs}\ and\ \citenamefont
  {Kothari}(2011)}]{Andrew2011}%
  \BibitemOpen
  \bibfield  {author} {\bibinfo {author} {\bibfnamefont {Andrew~M.}\
  \bibnamefont {Childs}}\ and\ \bibinfo {author} {\bibfnamefont {Robin}\
  \bibnamefont {Kothari}},\ }\bibfield  {title} {\enquote {\bibinfo {title}
  {Simulating sparse hamiltonians with star decompositions},}\ }in\ \href
  {\doibase 10.1007/978-3-642-18073-6_8} {\emph {\bibinfo {booktitle} {Theory
  of Quantum Computation, Communication, and Cryptography}}},\ \bibinfo
  {series} {Lecture Notes in Computer Science}, Vol.\ \bibinfo {volume} {6519}\
  (\bibinfo  {publisher} {Springer Berlin Heidelberg},\ \bibinfo {year}
  {2011})\ pp.\ \bibinfo {pages} {94--103}\BibitemShut {NoStop}%
\bibitem [{\citenamefont {Trotter}(1959)}]{Trotter1959product}%
  \BibitemOpen
  \bibfield  {author} {\bibinfo {author} {\bibfnamefont {Hale~F}\ \bibnamefont
  {Trotter}},\ }\bibfield  {title} {\enquote {\bibinfo {title} {On the product
  of semi-groups of operators},}\ }\href@noop {} {\bibfield  {journal}
  {\bibinfo  {journal} {Proceedings of the American Mathematical Society}\
  }\textbf {\bibinfo {volume} {10}},\ \bibinfo {pages} {545--551} (\bibinfo
  {year} {1959})}\BibitemShut {NoStop}%
\bibitem [{\citenamefont {Suzuki}(1976)}]{Suzuki1976generalized}%
  \BibitemOpen
  \bibfield  {author} {\bibinfo {author} {\bibfnamefont {Masuo}\ \bibnamefont
  {Suzuki}},\ }\bibfield  {title} {\enquote {\bibinfo {title} {Generalized
  trotter's formula and systematic approximants of exponential operators and
  inner derivations with applications to many-body problems},}\ }\href@noop {}
  {\bibfield  {journal} {\bibinfo  {journal} {Communications in Mathematical
  Physics}\ }\textbf {\bibinfo {volume} {51}},\ \bibinfo {pages} {183--190}
  (\bibinfo {year} {1976})}\BibitemShut {NoStop}%
\bibitem [{\citenamefont {Poulin}\ \emph {et~al.}(2015)\citenamefont {Poulin},
  \citenamefont {Hastings}, \citenamefont {Wecker}, \citenamefont {Wiebe},
  \citenamefont {Doherty},\ and\ \citenamefont {Troyer}}]{Poulin2015trotter}%
  \BibitemOpen
  \bibfield  {author} {\bibinfo {author} {\bibfnamefont {David}\ \bibnamefont
  {Poulin}}, \bibinfo {author} {\bibfnamefont {MATTHEW~B}\ \bibnamefont
  {Hastings}}, \bibinfo {author} {\bibfnamefont {Dave}\ \bibnamefont {Wecker}},
  \bibinfo {author} {\bibfnamefont {Nathan}\ \bibnamefont {Wiebe}}, \bibinfo
  {author} {\bibfnamefont {Andrew~C}\ \bibnamefont {Doherty}}, \ and\ \bibinfo
  {author} {\bibfnamefont {Matthias}\ \bibnamefont {Troyer}},\ }\bibfield
  {title} {\enquote {\bibinfo {title} {The trotter step size required for
  accurate quantum simulation of quantum chemistry},}\ }\href@noop {}
  {\bibfield  {journal} {\bibinfo  {journal} {Quantum Information \&
  Computation}\ }\textbf {\bibinfo {volume} {15}},\ \bibinfo {pages} {361--384}
  (\bibinfo {year} {2015})}\BibitemShut {NoStop}%
\bibitem [{\citenamefont {Daskin}\ \emph {et~al.}(2012)\citenamefont {Daskin},
  \citenamefont {Grama}, \citenamefont {Kollias},\ and\ \citenamefont
  {Kais}}]{Daskin2012universal}%
  \BibitemOpen
  \bibfield  {author} {\bibinfo {author} {\bibfnamefont {Anmer}\ \bibnamefont
  {Daskin}}, \bibinfo {author} {\bibfnamefont {Ananth}\ \bibnamefont {Grama}},
  \bibinfo {author} {\bibfnamefont {Giorgos}\ \bibnamefont {Kollias}}, \ and\
  \bibinfo {author} {\bibfnamefont {Sabre}\ \bibnamefont {Kais}},\ }\bibfield
  {title} {\enquote {\bibinfo {title} {Universal programmable quantum circuit
  schemes to emulate an operator},}\ }\href@noop {} {\bibfield  {journal}
  {\bibinfo  {journal} {The Journal of chemical physics}\ }\textbf {\bibinfo
  {volume} {137}},\ \bibinfo {pages} {234112} (\bibinfo {year}
  {2012})}\BibitemShut {NoStop}%
\bibitem [{\citenamefont {Daskin}\ \emph
  {et~al.}(2014{\natexlab{b}})\citenamefont {Daskin}, \citenamefont {Grama},\
  and\ \citenamefont {Kais}}]{Daskin2014universal}%
  \BibitemOpen
  \bibfield  {author} {\bibinfo {author} {\bibfnamefont {Anmer}\ \bibnamefont
  {Daskin}}, \bibinfo {author} {\bibfnamefont {Ananth}\ \bibnamefont {Grama}},
  \ and\ \bibinfo {author} {\bibfnamefont {Sabre}\ \bibnamefont {Kais}},\
  }\bibfield  {title} {\enquote {\bibinfo {title} {A universal quantum circuit
  scheme for finding complex eigenvalues},}\ }\href@noop {} {\bibfield
  {journal} {\bibinfo  {journal} {Quantum information processing}\ }\textbf
  {\bibinfo {volume} {13}},\ \bibinfo {pages} {333--353} (\bibinfo {year}
  {2014}{\natexlab{b}})}\BibitemShut {NoStop}%
\end{thebibliography}%

\end{document}